\documentclass[final,3p,times,twocolumn]{elsarticle}
\usepackage{blindtext, graphicx, amsmath, algorithm, algpseudocode, pifont, comment, layout, amsthm, amssymb}
\usepackage{enumitem}   
\usepackage{makecell}
\usepackage{float}
\usepackage{balance}
\usepackage{amsmath,amssymb,amsfonts, amsthm}
\usepackage{algorithm}
\usepackage{bbm}
\usepackage{graphicx}
\usepackage{textcomp}
\usepackage{xcolor}
\usepackage{kotex}
\usepackage{subfigure}
\usepackage{booktabs}
\usepackage{multicol}
\usepackage{multirow}
\usepackage[normalem]{ulem}
\usepackage{lettrine}

\usepackage{xspace}

\newcommand{\BfPara}[1]{{\noindent\bf#1.}\xspace}

\usepackage[utf8]{inputenc}
\usepackage[english]{babel}
\usepackage{kotex} 
\usepackage{caption}
\journal{ICT Express}

\begin{document}
\begin{frontmatter}    
\title{Software Simulation and Visualization of Quantum Multi-Drone Reinforcement Learning}
\author{$^{\dag}$Chanyoung Park} 
\ead{cosdeneb@korea.ac.kr}
\author{$^{\dag}$Jae Pyoung Kim}
\ead{paulkim436@korea.ac.kr}
\author{$^{\dag}$Won Joon Yun}
\ead{ywjoon95@korea.ac.kr}
\author{$^{\dag}$Soohyun Park\corref{cor1}}
\ead{soohyun828@korea.ac.kr}
\author{$^{\S}$Soyi Jung\corref{cor1}}
\ead{sjung@ajou.ac.kr}
\author{$^{\dag}$Joongheon Kim}
\ead{joongheon@korea.ac.kr}
\address{$^{\dag}$School of Electrical Engineering, Korea University, Seoul, Republic of Korea
\\
$^{\S}$  Department of Electrical and Computer Engineering, Ajou University, Suwon, Republic of Korea
}
\cortext[cor1]{Corresponding authors}
\fntext[eq2]{\noindent This research was funded by the National Research Foundation of Korea (2022R1A2C2004869). Chanyoung Park and Jae Pyoung Kim contributed equally (first authors).}
\begin{abstract}
Quantum machine learning (QML) has received a lot of attention according to its light training parameter numbers and speeds; and the advances of QML lead to active research on quantum multi-agent reinforcement learning (QMARL). Existing classical multi-agent reinforcement learning (MARL) features non-stationarity and uncertain properties. Therefore, this paper presents a simulation software framework for novel QMARL to control autonomous multi-drones, i.e., quantum multi-drone reinforcement learning. Our proposed framework accomplishes reasonable reward convergence and service quality performance with fewer trainable parameters. Furthermore, it shows more stable training results. Lastly, our proposed software allows us to analyze the training process and results.\end{abstract}


\end{frontmatter}

\section{Introduction}
\BfPara{Background and Motivation} 
Spurred by the recent advances in quantum computing (QC), attempts to re-implement existing machine learning (ML) have been presented to discover quantum advantages~\cite{KWAK2022,jerbi2021parametrized}. Primarily, multi-agent reinforcement learning (MARL) is one of the most challenging ML fields due to scalability and non-stationarity~\cite{yun2022quantum}. Yun \textit{et al.} have proposed QMARL that can perform reasonably by using a small number of trainable parameters compared to classical neural networks~\cite{yun2022quantum}. 
In addition, meta QMARL has been proposed, which enables adaptation to context changes by using a tiny size of memory (known as pole memory in QC) compared to the classical method~\cite{yun2022quantum}. Although these studies have shown the effectiveness and feasibility, user-friendly simulation software is not considered and designed yet, whereas the software simulator implementation and visualization research is considered as one of key topics in network and mobility research fields~\cite{tvt202006kamel,sim2,sim3}.

For existing classical RL, OpenAI has already developed and publicized Gym~\cite{brockman2016openai}, which helps compare the performance between various RL algorithms by implementing communication between the algorithms and the environments. Furthermore, it has become standard to evaluate RL algorithms via Gym environments. Many versions of Gym exist for different frameworks of RL. For example, Gazebo is a library for 3D robotics simulation in which RL can be utilized. Petting Zoo~\cite{terry2020pettingzoo} and Starcraft multi-agent challenge (SMAC)~\cite{samvelyan19smac} are both environments for MARL, equipped with multiple agents. 
Next, for QML frameworks, a variety of tools such as \textit{TorchQuantum}~\cite{quantumnas}, and \textit{Qiskit}~\cite{aleksandrowicz2019qiskit} has also been developed and publicized in recent years. These libraries significantly contributed to the research of many current QML algorithms. For example, the QMARL framework mentioned above~\cite{yun2022quantum} also utilizes \textit{TorchQuantum} for the implementation.

Similar to the frameworks introduced above, we have designed a QMARL software framework for visualizing a multi-drone communication environment~\cite{tvt201905shin,jcn2022lee,icc,tvt202106jung}, so-called quantum multi-drone reinforcement learning (QMDRL). The MDRL framework proposed in this paper is a hybrid model which utilizes both classic and QC. Classical computing is used to implement the \textsf{Trainer} in Fig.~\ref{fig:Sys architure}, which computes the loss function, performs optimization via gradient descent, and updates the target values. QC is used to implement the quantum-based policy (Q-policy) in \textsf{Q-policy Layer}, which computes the action distribution of all the drone agents. In this process, the data re-uploading technique encodes both classical data and compute output data. Leveraging QC in our proposed model has not only increased the computational speed exponentially but also shown a higher total reward value after training. Therefore, the proposed model outperforms the previous model and other benchmarks, corroborating our methodology's efficiency. Moreover, we can analyze multi-agent training processes and results thanks to our proposed visual simulation software.

\BfPara{Contributions} The contributions of this work are as follows: 
(i) A proposed QMDRL framework for a multi-drone environment is designed to maximize users' support rate and quality of service (QoS). Based on the proposed framework, we construct visualized human-computer/drone interaction (HCI) interface for better understanding to the system designers and engineers;   (ii) We validate that our proposed QMDRL training algorithm has more stable and effective performances than the classical MARL based on our proposed visualization software framework; 
(iii) A Q-policy network that computes drone agents' action distribution via data re-uploading is proposed. The data re-uploading scheme helps our Q-Policy to stably handle multiple qubits.
   
\BfPara{Organization} 
The rest of this paper is organized as follows.
Sec.~\ref{sec:2} introduces the basic concepts of quantum machine learning. 
Sec.~\ref{sec:3} proposes a software simulator/tool for quantum multi-drone reinforcement learning. 
Sec.~\ref{sec:4} evaluates and demonstrates the software simulator of quantum multi-drone reinforcement learning. 
Sec.~\ref{sec:5} concludes this paper.

\section{Basics of Quantum Computing}\label{sec:2}


\begin{figure*}[ht]
    \centering
    \includegraphics[width=0.990\linewidth]{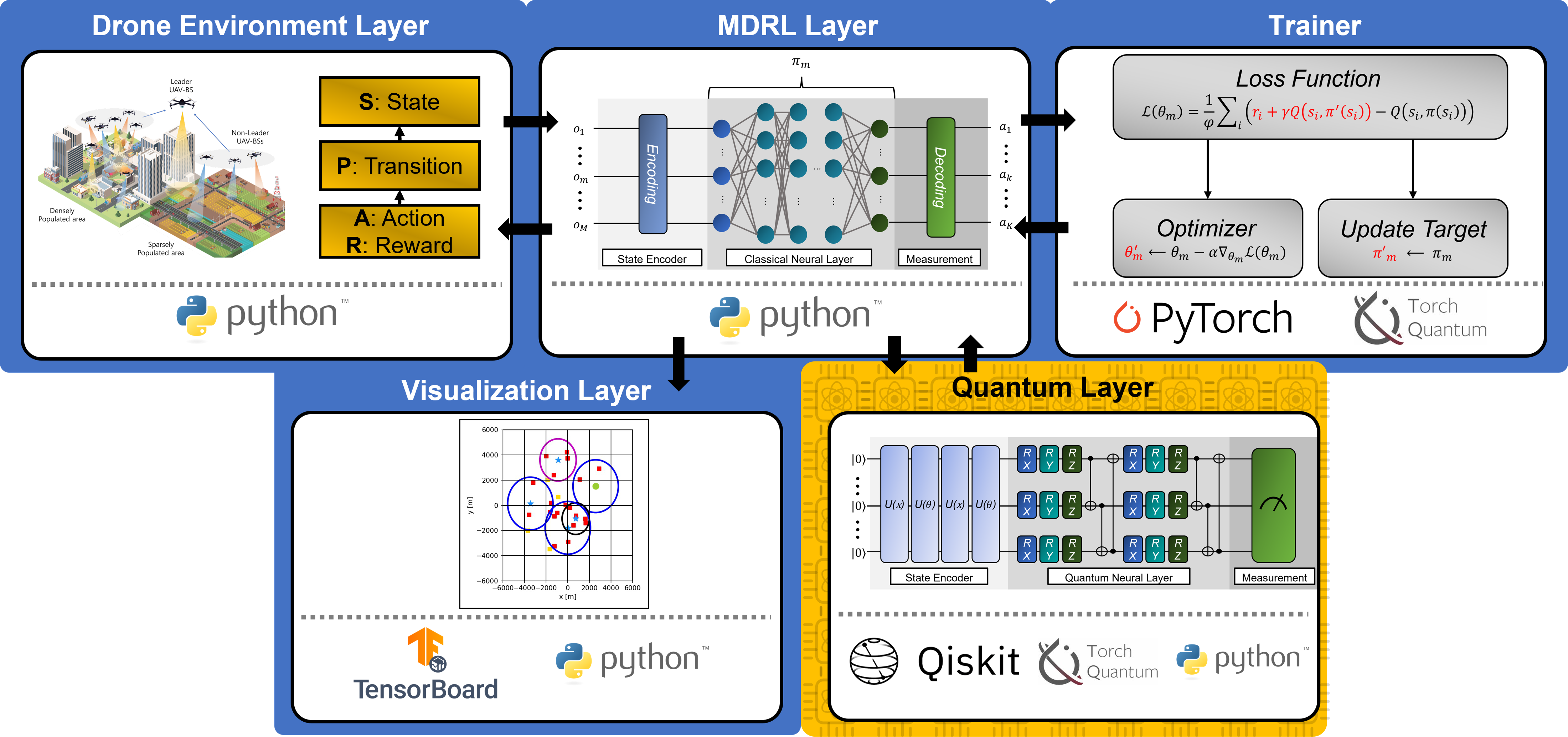}
    \caption{The illustration of visual simulation software for quantum multi-drone reinforcement learning. The \textit{blue} and \textit{yellow} blocks indicate that each block utilizes classical computing, and quantum computing, respectively.}
    \label{fig:Sys architure}
\end{figure*}

\BfPara{Qubits}
In QC, a \textit{qubit} is used as the basic unit of information instead of bits. While bits can only be expressed by 0 or 1, qubits can be expressed as a combination of two bases, $|0\rangle$ and $|1\rangle$. This nature is also known as \textit{superposition}~\cite{bouwmeester2000physics}. Furthermore, quantum states can be expressed as qubits, and entanglement can occur between qubits which causes individual qubits to be strongly correlated. Due to these differences, qubits can express more information than classical bits. 
Assuming there is a $q$ qubits system, the quantum state defined within the Hilbert state can be defined as follows, 
$|\psi\rangle=\alpha_{1}|0\cdots 0\rangle + \cdots + \alpha_{2^q}|1\cdots 1\rangle$,
where $\alpha$ represents the probability amplitude and $\sum^{2^q}_{i=1}\alpha_{i}^2=1$; finally, the quantum state can be graphically expressed on the Bloch sphere.

\BfPara{Basic Quantum Gates}
Although bits and qubits are different, classical data can be encoded into qubits via quantum state encoders composed of basic quantum rotation gates expressed as $R_x$, $R_y$, and $R_z$. These rotation gates encode classical bits into qubits and perform unitary operations on a qubit by rotating it in the direction of $x$-, $y$-, and $z$- axes, respectively.
Additionally, as mentioned above, qubit entanglement between two qubits can be achieved by a \textit{CNOT} gate. This is done by performing the \textit{XOR} operation on one qubit and using the other as the control qubit. All the gates introduced above are used to compose quantum neural networks.

\BfPara{Quantum Neural Networks}
The structure of quantum neural networks (QNN) comprises the state encoder, parameterized quantum circuit (PQC), and the measurement layer. Since classical data $X$ is incompatible with quantum circuits, it must be converted into quantum states before being used as input data. In this paper, the data re-uploading~\cite{P_rez_Salinas_2020} method will be used in implementing the state encoder as shown in Fig. \ref{fig:Sys architure}. Data re-uploading and quantum state encoding are jointly achieved by passing $q$ number of $|0\rangle$ qubits through a sequence of unitary operation gates containing the information of $X$ and trainable parameters of the encoder, denoted as $\theta_{enc}$. In order to imbue each gate with information of $X$, it has to be split into $[x_{1} \cdots x_{N}]$. Then, these gates will repeatedly rotate the initial qubits, and the output quantum states produced successfully convert $X$ into quantum states. The process of state encoding can be expressed as follows,
$|\psi_{\text{enc}}\rangle = U({\theta}_{N})U(\mathbf{x}_{N})\cdots U({\theta}_{1})U(\mathbf{x}_{1}) |\psi_{0}\rangle$,
where $N$ is the number of split data, $U(\cdot)$ represents a unitary operation of qubit rotation, $|\psi_0\rangle$ is the initial quantum state, $|\psi_{enc}\rangle$ is the encoded qubit and $\theta$ is the trainable parameters of the state encoder. 
The PQC will then process the converted qubits. As shown in Fig. \ref{fig:Sys architure}, the PQC is composed of the $R_x$, $R_y$, $R_z$ and the \textit{CNOT} gates which contain trainable parameters $\theta_{PQC}$. By rotating and entangling the input qubits, we will be able to obtain the values needed for MARL. This can be expressed as $|\psi_{PQC}\rangle=U(\theta_{PQC})|\psi_{\text{enc}}\rangle$.
Finally, the produced qubits will proceed into the measurement layer, where they will be measured. Measurement can be carried out by applying a projection matrix $M \in \textbf{M} \equiv \{M_{1},\cdots,M_{c},\cdots,M_{C}\}$ onto the $z-$ axis. The value produced from this operation is known as \textit{observable}, denoted as $\langle V \rangle_{\theta}\in [-1,1]^{C}$.
The measurement operation can be expressed as follows,
    $\langle V_{c}\rangle_{\theta}=\langle 0|U^{\dagger}(x)U^{\dagger}(\theta)M_{c}U(\theta)U(x)|0\rangle = \langle \psi|M_{c}|\psi\rangle$,
where $(\cdot)^{\dagger}$ is the complex conjugate operator. The \textit{observable} produced from the quantum layer is the action distribution of the actors and is used to calculate the loss function. However, the function cannot be trained by backpropagation because quantum states within the QNN cannot be measured and will collapse when the chain rule is applied. Therefore, training is carried out by computing the loss gradient from the symmetric difference quotient of the loss function.

\section{Software Simulator for Quantum Multi-Drone Reinforcement Learning}\label{sec:3}

\begin{figure}[!t]
\centering 
\begin{tabular}{c}
\includegraphics[width=0.99\columnwidth]{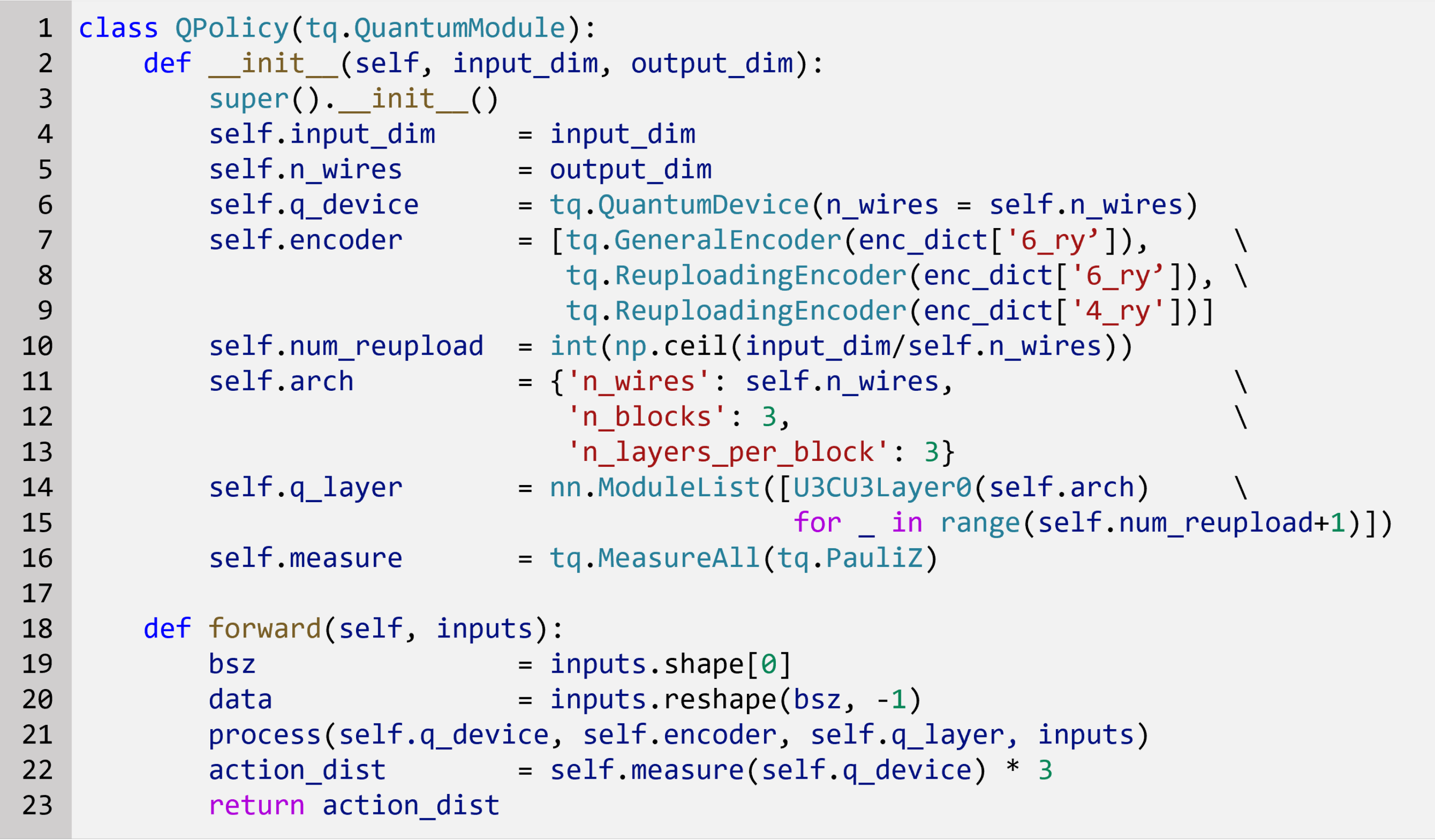}\\
(a) Policy file (policy.py).\\
\includegraphics[width=0.99\columnwidth]{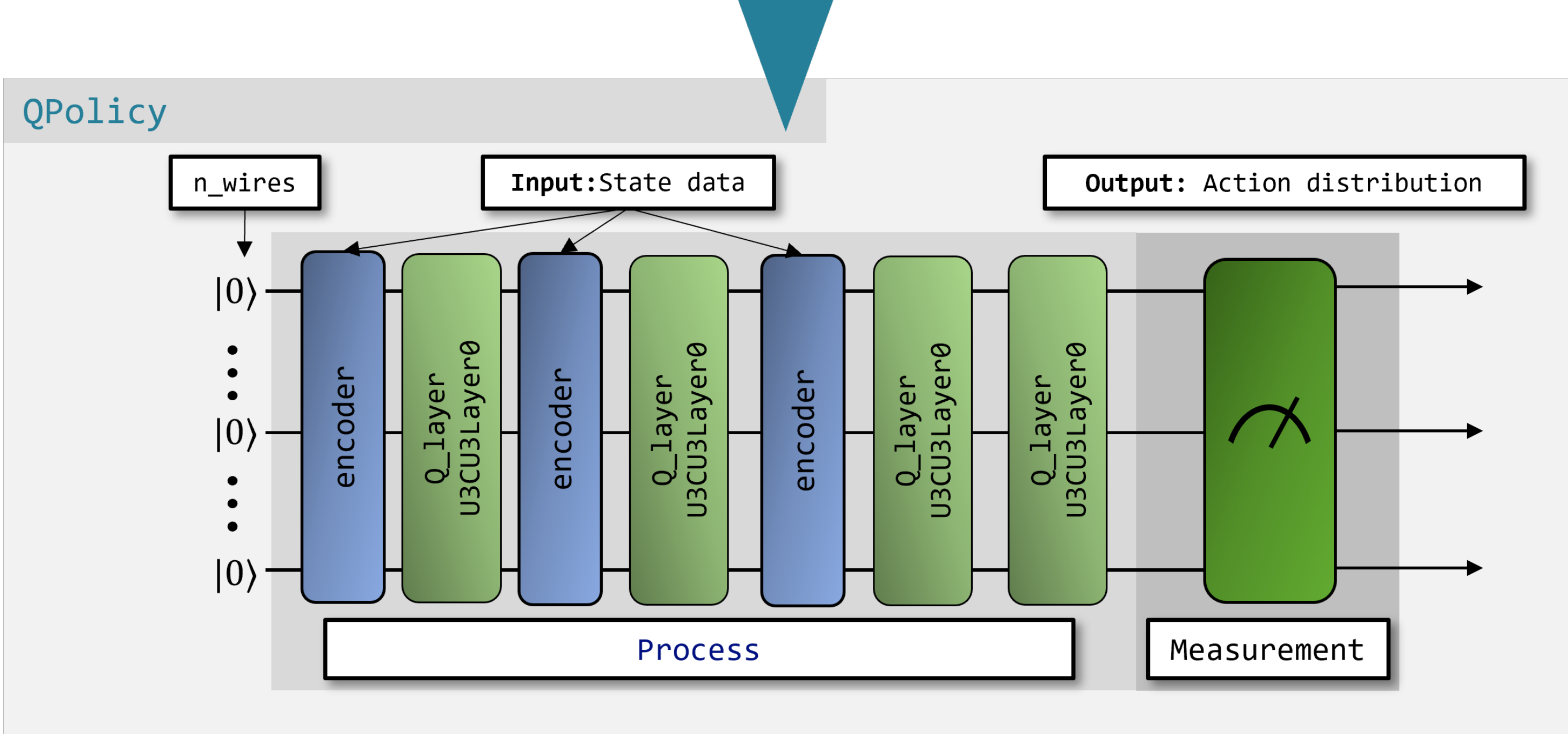}\\
(b) The illustration of Q-policy.
\end{tabular}
\caption{Quantum Layer.}
\label{fig:Qpolicy}
\end{figure}

\begin{figure}[!t]
\centering 
\begin{tabular}{c}
\includegraphics[width=0.99\columnwidth]{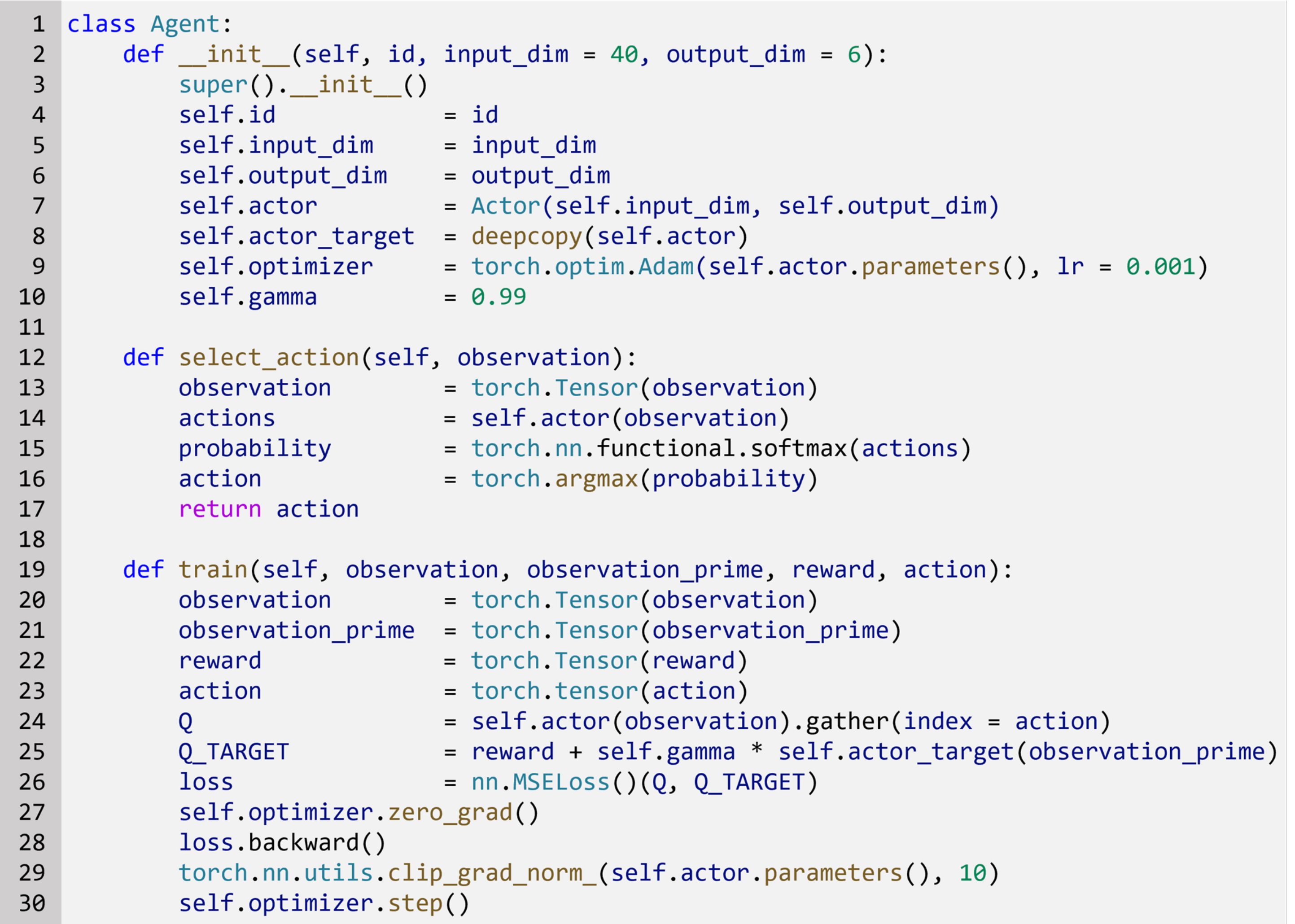}\\
\end{tabular}
\caption{Drone Agent's Training Process.}
\label{fig:training process}
\end{figure}

\begin{figure}[!ht]
    \centering
    \includegraphics[width=0.9885\columnwidth]{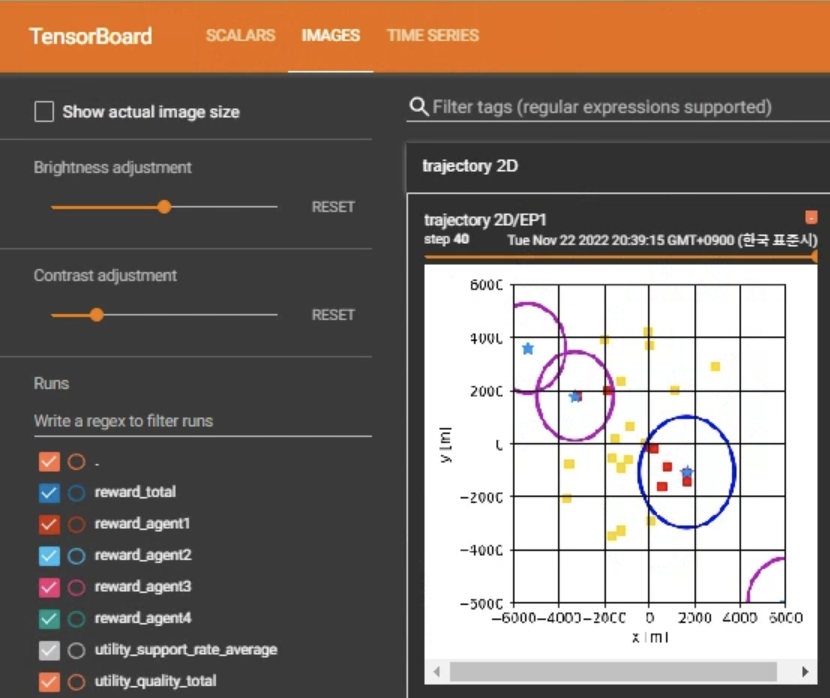}
    \caption{Visualization with \textit{tensorboard}.}
    \label{fig:tensorboard}
\end{figure}

\BfPara{Framework of Proposed QMARL}
We consider the MARL environment in~\cite{park2022cooperative}, where drone agents try to learn a policy that maximizes the total reward. We describe the sequential framework of our proposed QMDRL based on Fig.~\ref{fig:Sys architure}. 
\begin{enumerate}
    \item In \textsf{Drone Environment Layer}~\cite{park2022cooperative}, all $M$ drone agents collect observation information denoted as the set of states, which is $\textbf{S}\equiv[o_1,\cdots ,o_m,\cdots,o_{M}]$. 
    \item With drone agents' state, they take actions in time step $t$ based on their policy (\textit{i.e.}, actor), and then the state is transitioned to next time step $t+1$ in \textsf{MDRL Layer}. Here, all actions and rewards that drone agents take are denoted as the following sets, $\textbf{A}\equiv[a_1,\cdots ,a_m,\cdots,a_{M}]$ and $\textbf{R}\equiv[r_1,\cdots ,r_m,\cdots,r_{M}]$. 
    \item Loss value is calculated by criticizing the reward that drone agents get at time step $t$ with the return by target actor network in \textsf{Trainer Layer}. After that, the optimizer updates the parameters of actor networks in the direction of decreasing the value of the loss function. In the case of the target network, its parameters are updated at a specific time intermittently. The detailed training processes are described in Sec.~\ref{sec:training process}. 
    \item The \textsf{Quantum Layer}, also known as the Q-policy network, produces the action distribution of all the drone agents by taking the state data as input. The output data is computed by the encoders and parameterized quantum circuits which are the components of the Q-policy network. Then, the action distribution values will be used to calculate the loss function utilized in the trainer layer via the mean squared error function. Finally, the gradient of the loss function is calculated to perform gradient descent to update the actor and the target. 
    \item At the same time, all drone agent's trajectories are visualized while training their policies by \textit{tensorboard} in the \textsf{Visualization Layer} as shown in Fig.~\ref{fig:tensorboard}.
\end{enumerate}

\begin{figure*}
    \centering
    \subfigure[Cumulative Reward.]
    {
    \includegraphics[width=0.30\linewidth]{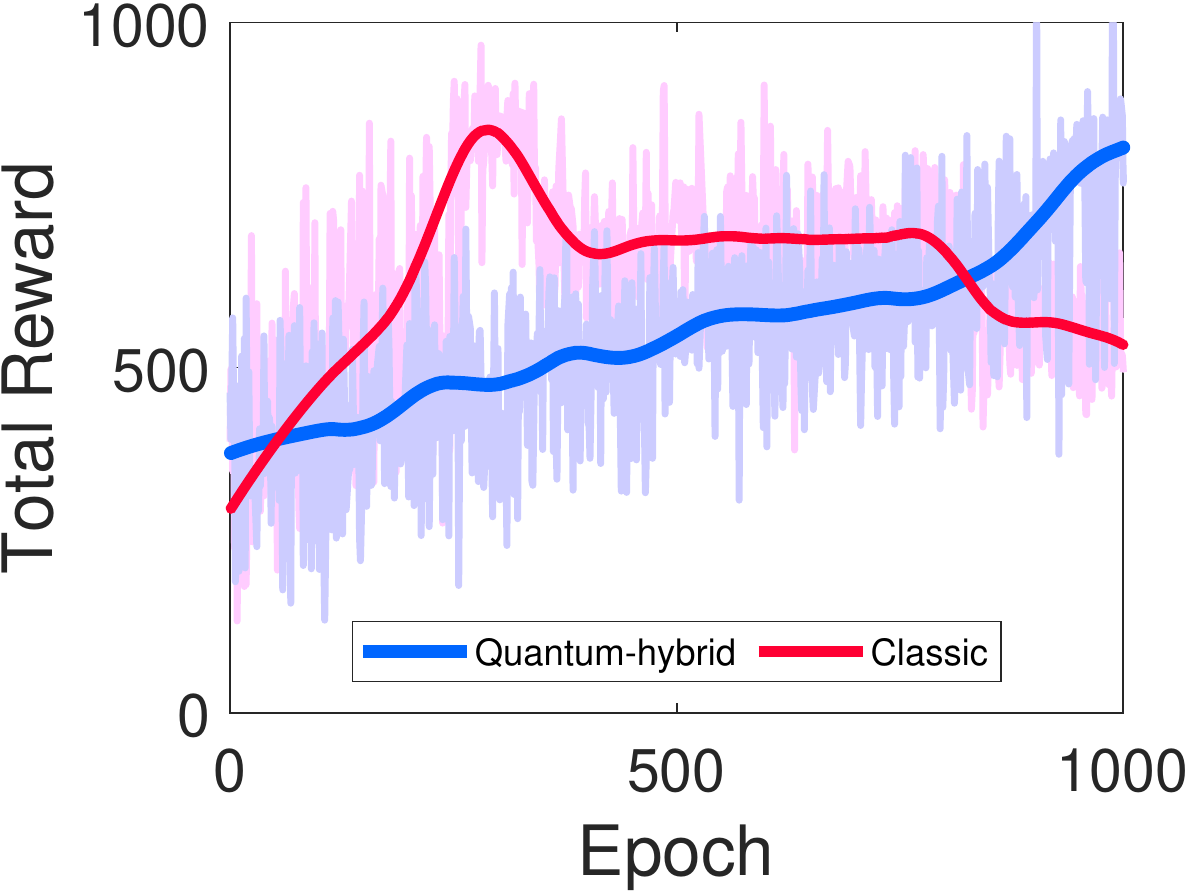}
    \label{fig:Reward}
    }
    \subfigure[Support Rate.]
    {
    \includegraphics[width=0.30\linewidth]{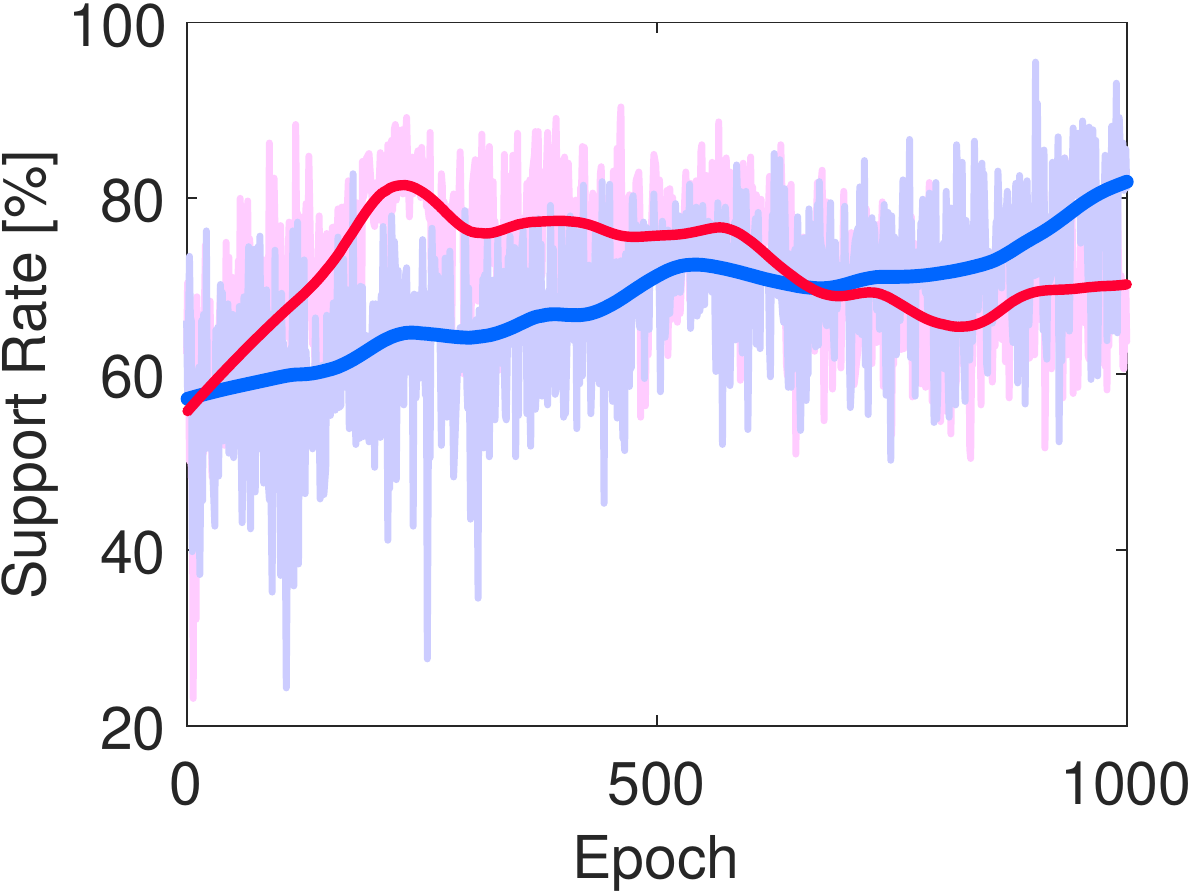}
    \label{fig:Support_rate}
    }
    \subfigure[Quality of Service.]
    {
    \includegraphics[width=0.30\linewidth]{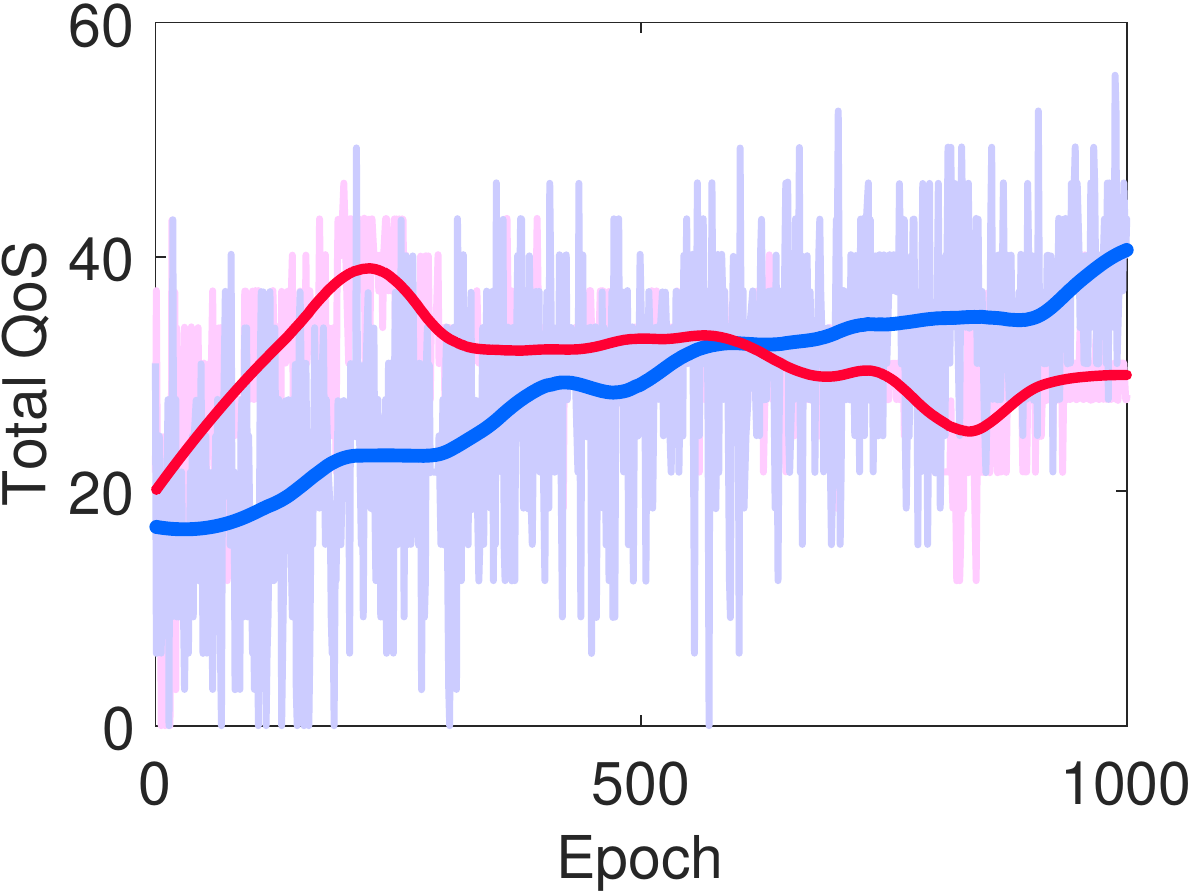}
    \label{fig:QoS}
    }
    \caption{Various aspects of results for evaluating performance in each training method. The \textit{blue} and \textit{red} represent the learning results in our proposed QMDRL and classical MARL, respectively.}
\end{figure*}

\begin{figure*}[t!]
    \centering
    \includegraphics[width=0.99\columnwidth]{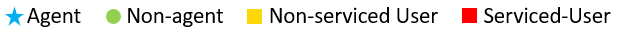}\\
    \subfigure[$t = 0$.]
    {
    \includegraphics[width=0.376\columnwidth]{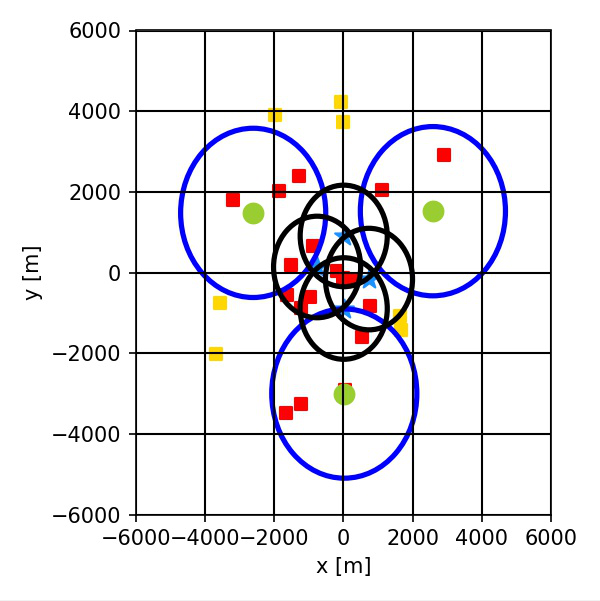}
    \label{fig:t=0}
    }
    \subfigure[$t = 5$.]
    {
    \includegraphics[width=0.376\columnwidth]{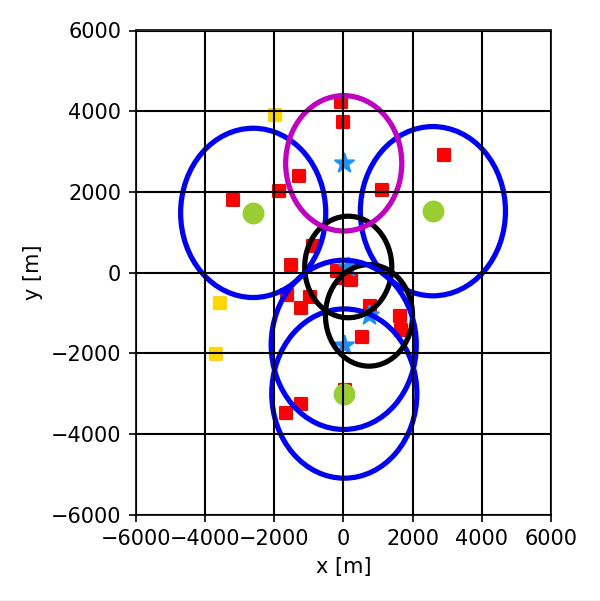}
    \label{fig:t=5}
    }
    \subfigure[$t = 10$.]
    {
    \includegraphics[width=0.376\columnwidth]{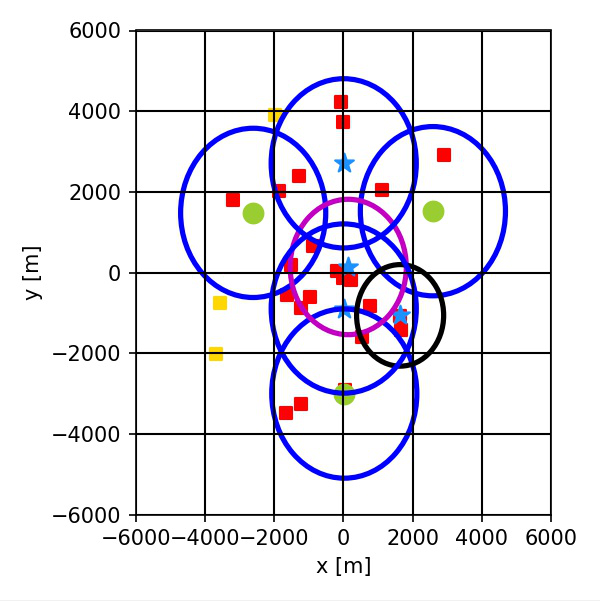}
    \label{fig:t=10}
    }
    \subfigure[$t = 15$.]
    {
    \includegraphics[width=0.376\columnwidth]{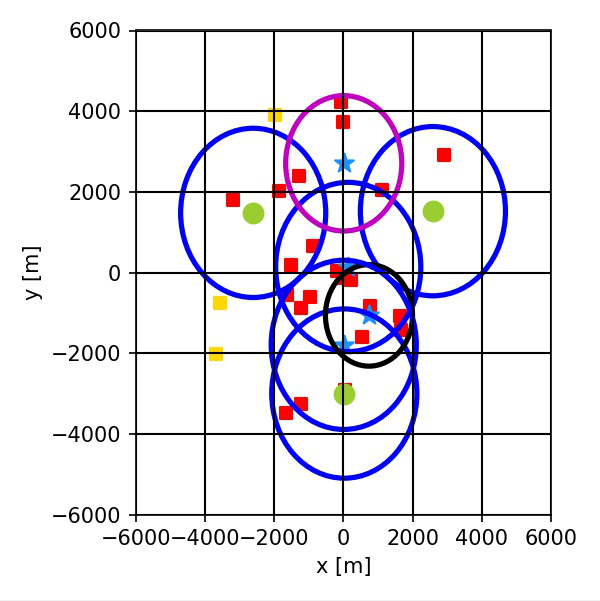}
    \label{fig:t=15}
    }\\
    \subfigure[$t = 20$.]
    {
    \includegraphics[width=0.376\columnwidth]{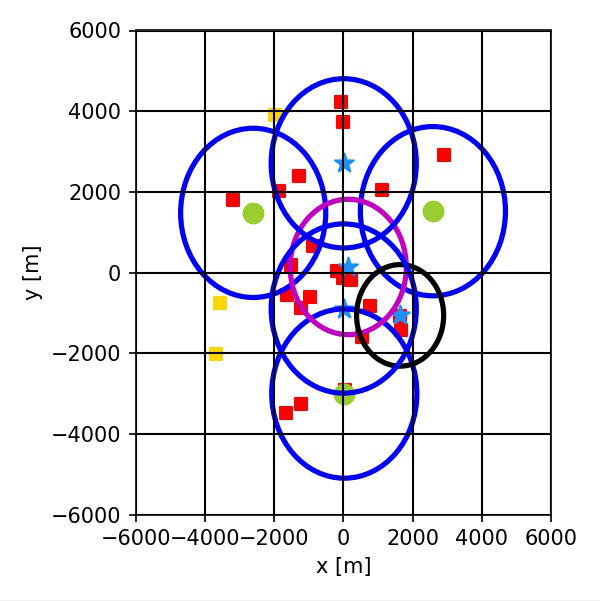}
    \label{fig:t=20}
    }
    \subfigure[$t = 25$.]
    {
    \includegraphics[width=0.376\columnwidth]{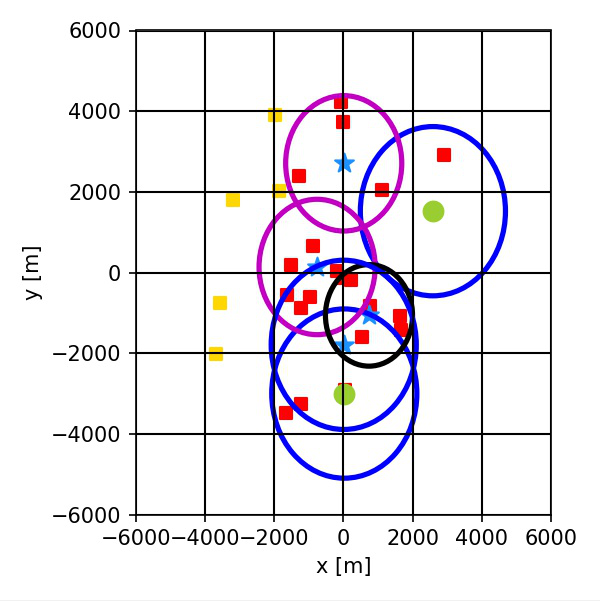}
    \label{fig:t=25}
    }
    \subfigure[$t = 30$.]
    {
    \includegraphics[width=0.376\columnwidth]{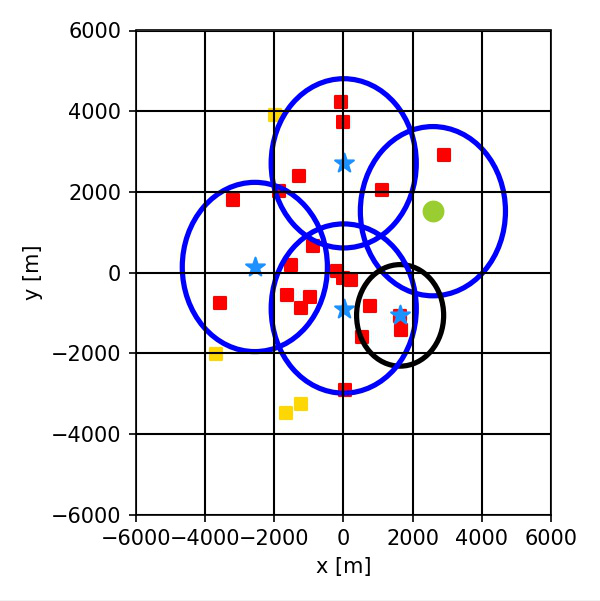}
    \label{fig:t=30}
    }
    \subfigure[$t = 35$.]
    {
    \includegraphics[width=0.376\columnwidth]{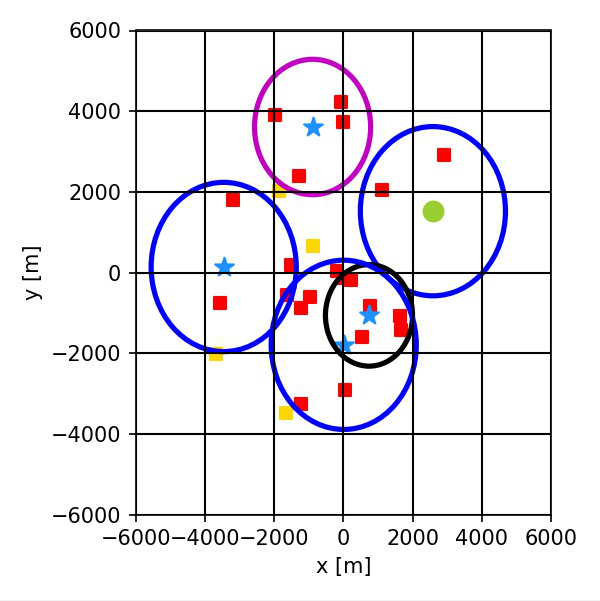}
    \label{fig:t=35}
    }
    \subfigure[$t = 40$.]
    {
    \includegraphics[width=0.376\columnwidth]{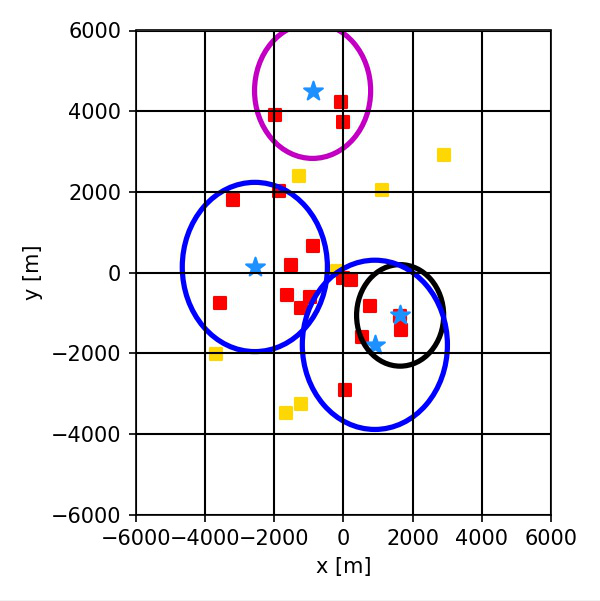}
    \label{fig:t=40}
    }\\
    \caption{Trajectories of drones trained by the proposed QMDRL.}
\end{figure*}

\BfPara{Description of Q-policy}
Each drone agent has our proposed Q-policy as shown in Fig.~\ref{fig:Qpolicy}.
Especially, the algorithm for the Q-policy network is as shown in Fig. \ref{fig:Qpolicy}(a). 
Firstly, (Lines 4--5) initializes the input dimensions as the size of the observation data of the drone agents and the initial quantum state dimension as the output size equivalent to the total number of actions. 
Next, (Line 6) defines the quantum device where the encoders and quantum layers will be placed.     
(Lines 7--9) and (Line 14) define the encoder and quantum layer, respectively. As shown in the figure, the encoder is composed of the pre-defined general encoder and re-uploading encoder, while the quantum layer is made up of \textit{Controlled U3} gates.     
In (Line 10), the number of repetitions of data re-uploading is set as the ceiling value of the input dimension divided by the number of initial qubits.     
(Lines 11--13) defines the parameter values of the \textit{Controlled U3} gate.     
Finally, (Line 16) defines the measurement layer as applying \textit{Pauli-Z} gate to all the qubits within the circuit. Fig. \ref{fig:Qpolicy}(a) contains the code for the forward function of the Q-policy network. 
(Lines 19--20) shows the process of reshaping the input data into a single-column matrix.
Then, (Line 21) is the part where data re-uploading occurs. The process is repeated for the number of repetitions initialized above. Every iteration passes the input data through one encoder and one quantum-layer (Q-layer). Note that the first and the last iterations use encoders with different input dimensions. 
Lastly, (Lines 22--23) returns the action distribution by measuring the output produced by data re-uploading.

Fig. \ref{fig:Qpolicy} also shows the structure of the Q-policy network, which is designed to perform data re-uploading. As shown in the figure, data re-uploading passes an initial quantum state through a repeated sequence of encoders and Q-layers. According to the algorithm elaborated above, the quantum gates in the encoders are imbued with the input state data. At the same time, the Q-layer is made up of \textit{Controlled U3} gates containing parameters such as the number of blocks, the number of layers per block, and the number of qubits. After data re-uploading has been performed, the output quantum state is measured by projecting the quantum state onto the reference $z$-axis to obtain the action distribution data.

\BfPara{Training Process of Drone Agents}
\label{sec:training process}
Fig.~\ref{fig:training process} shows the training process of drone agents in our MARL environment with their policies. 
(Lines 2--9) performs the initialization of the policy's parameters, where each drone agent is given a unique identification to distinguish drone agents in (Line 3).
(Lines 7--8) creates two neural networks corresponding to the drone agent's policy; actor-network and target actor-network. Using the target network, the learning process becomes more stable by making the target value independent to the training parameters in actor-network.
(Line 13) converts the drone agent's observation into the hidden state by encoding the parameters of the model machined to be suitable to use \textit{PyTorch}-related tools (\textit{i.e., TorchQuantum} and  \textit{Qiskit}). 
After that, when the actor-network object is called with a hidden state in (Line 14), the drone agent executes the \textsf{forward} function to return the action set of the drone. 
In (Lines 15--17), \textit{softmax} in \textit{PyTorch} is a function that outputs probabilities of all actions, and the drone agent returns the action with the highest probability.
In RL, the drone agent's training is performed after state transition, where the transition pair $\left(\textbf{S}, \textbf{A}, \textbf{R}, \textbf{S}'\right)$ is encoded into hidden states in (Lines 20--23). Denotation $'$ is to distinguish whether the set is for $t+1$. 
(Lines 24--25) calculates the Q-function of actor-network and target actor-network.
In (Lines 27--30), the optimizer updates the parameters of actor networks to decrease the value of the loss function by \textsf{Trainer} in Fig.~\ref{fig:Sys architure}.

\section{Performance Evaluation}\label{sec:4}



\BfPara{Visualization of Drone Agents' Training Results}
Thanks to our visualization using \textit{tensorboard}, we can analyze drone agents' decisions over time. Fig.~\ref{fig:Reward}--(c) show the total reward drone agents get, the support rate among all users, and users' QoS in all epochs, respectively. All results show the common tendencies, where drone agents in the classical MARL get larger values in the intermediate step of learning and a faster increase rate in the early step of learning. However, users in the QMDRL finally get more considerable value at the end of the training. In other words, our proposed QMDRL shows a more stable convergence rate with little fluctuations and higher values at the end of the learning. We can also observe drone agents' trajectories while learning their policies in Fig.~\ref{fig:t=0}--(i), where all drone agents try to provide high-quality wireless communication service to as many people as possible. When there is no malfunctioning, drone agents move to areas where any drone doesn't provide service to users. Conversely, when any malfunction occurs, the drone agents move to areas where the malfunctioned non-drone/agent has provided service to users. By our proposed visual simulation software framework, we can validate that the QMDRL can make a reasonable performance with a smaller number of trainable parameters compared to classical MARL in reward convergence and system quality.

\BfPara{Demonstration Video} 
The video demonstration for our proposed visual simulation software framework and simulation results are in~\cite{demovideolink}. 
First of all, this demo shows the architecture of our Q-Policy network, and it proposes a visualization framework of software simulation for QMDRL visualization. After that, we show the HCI process sequentially using our framework. 


\section{Concluding Remarks}\label{sec:5}
We present a novel QMDRL framework for MARL-based autonomous multi-drone mobility control/coordination and analyze the performances by visualization using \textit{tensorboard}. In addition, we show how our proposed QMDRL could be used to train the multi-drone/agent system efficiently. As future work, we plan to investigate other applications of QMDRL in versatile situations using our proposed visualization framework.

\bibliographystyle{elsarticle-num}
\bibliography{reference}
\end{document}